\documentclass{birkmult}
 \theoremstyle{definition}
 
 \theoremstyle{remark}

 \numberwithin{equation}{section}

\usepackage{color,tensor}

\begin{document}
%
%
%
%
%
%
%
%
%
\title[Spin Chains, Graphs and State Revival]
{Spin Chains, Graphs and State Revival}

\author{Hiroshi Miki}
\address{Meteorological College\\ Asahi-Cho\\ Kashiwa\\ 277 0852\\ Japan}
\email{hmiki@mc-jma.go.jp}

\author{Satoshi Tsujimoto}
\address{Department of Applied Mathematics and Physics\\ Graduate School of Informatics\\
Kyoto University\\ Sakyo-Ku\\ Kyoto\\ 606 8501\\ Japan}
\email{tsujimoto.satoshi.5s@kyoto-u.jp}

\author[Luc Vinet]{Luc Vinet}

\address{%
Centre de Recherches Math\'{e}matiques\\
Universit\'{e} de Montr\'{e}al\\
P.O. Box 6128\\
Centre-ville Station\\
Montr\'{e}al (Qu\'{e}bec)\\
H3C 3J7\\
Canada}

\email{vinet@crm.umontreal.ca}

\thanks{The research of ST is supported by JSPS
KAKENHI (Grant Numbers 16K13761) and that of LV by a discovery grant of the
Natural Sciences and Engineering Research Council (NSERC) of Canada.}


\date{}

\begin{abstract}
Connections between the 1-excitation dynamics of spin lattices and quantum walks on graphs will be surveyed. Attention will be paid to perfect state transfer (PST) and fractional revival (FR) as well as to the role played by orthogonal polynomials in the study of these phenomena. Included is a discussion of the ordered Hamming scheme, its relation to multivariate Krawtchouk polynomials of the Tratnik type, the exploration of quantum walks on graphs of this association scheme and their projection to spin lattices with PST and FR.
\end{abstract}

\maketitle
\section{Introduction}

The main objective of this lecture is to illustrate the role that orthogonal polynomials play in the analysis of spin network dynamics which are of relevance for quantum information. The transfer of quantum states and the generation of entangled states are two important tasks in this context. Let us first indicate broadly how we shall describe these mathematically in the following.\par
Take quite generally, a finite set of sites labelled by the integers $n=0, 1, \ldots, N.$ Let $\left| n\right>$  be the characteristic vector in $\mathbb{C}^{N+1}$ which has entry $1$ at position $n$ and zeros everywhere else. Consider this bra as the vector representing the quantum state of interest at the location $n$. The evolution operator is the unitary $U(t) = e^{-itH}$ with $t$ the time and $H$ some Hermitian Hamiltonian operator; therefore the state $\left| n\right>$ at time $t=0$ will become $U(t)\left| n\right>$ at time $t$.
Pick a reference site say $0$. We shall wish to have a dynamics such that for some $t=T$,
\begin{equation}
U(T)\left| 0\right>=\alpha \left| 0\right> +\beta \left| N\right> ,\quad |\alpha |^2+|\beta |^2=1.
\end{equation}
If this happens, we shall say that we have Fractional Revival (FR) at two sites. Two special cases are of particular interest:
\begin{itemize}
\item[] \textit{Perfect State Transfer (PST)} \\
If $\alpha = 0$ and $\beta = e^{i\phi }$, one has $U(T)\left| 0\right> = e^{i\phi} \left| N\right>$, that is the state $\left| 0\right>$ at time $t=0$ is found at time $t=T$ with probability $1$ at the site $N$; one thus say that it has been perfectly transferred.\\
\item[] \textit{Generation of maximal entanglement}\\
If the norm of both $\alpha $ and $\beta $ is equal to $\frac{1}{\sqrt{2}}$, the resulting state at $t=T$ is equivalent to $\frac{1}{\sqrt{2} }\left( \left| 0\right> + \left| N\right)\right>$ which can be pictured as the sum of two vectors: one with a spin up at site $0$ and spin down at all the other sites and the other with its only spin up at the site $N$. This is manifestly a maximally entangled state (that cannot be written as the product of two vectors). By evolving the vector $\left| 0\right> $ under $U(t)$ we have thus generated maximal entanglement at time $T$.
\end{itemize}
These two tasks can be achieved in properly engineered spin chains if one focuses on one-excitation dynamics \cite{Bo,BV,Ka,VZ1}. PST has in fact been modelled and experimentally realized in photonic waveguide arrays \cite{CS,PK}.\par
In the language of Graph Theory, such processes can be viewed as Quantum Walks on weighted paths (one-dimensional graphs) \cite{CFG,CDDEKL,Go,KT}. Quantum walks on other types of graphs have been used in the design of algorithms such as the ones of Grover or Ambianis for example \cite{CG,FG}. The interest in PST has prompted the examination of this process on spin networks deployed on graphs other than paths with the Hamiltonian taken to be the adjacency matrix (or the Laplacian).\par
The goals of this lecture are the following:
\begin{enumerate}
\item To describe situations where one-excitation dynamics in spin lattices with PST correspond to quantum walks on graphs.
\item To show that spin systems with PST can conversely be identified through this correspondence i.e. by projecting from graphs.
\item To consider for illustration, graphs that belong to the Hamming and generalized Hamming schemes. 
\item To show that the "go-betweens" are orthogonal polynomials of the Krawtchouk type in one or more variables. As a bonus, this will offer a review of their role in spectral graph theory.
\end{enumerate}\par
The outline of the lecture is as follows. In the next section, we shall discuss in some details how spin chains can be engineered so as to exhibit FR and/or PST. As shall be seen, this will involve conditions on the spectrum and the reconstruction of the Hamiltonians from these constrained data.  Orthogonal polynomials will be seen to play a central role in such inverse spectral problems. The spin chains associated to two families of orthogonal polynomials will be introduced as examples in Section 3. This will give the occasion to review some properties of the Krawtchouk polynomials as well as of the novel para-Krawtchouk polynomials. It will be noted that this last family stems naturally from the exploration of fractional revival. In Section 4, after an elementary review of the binary Hamming scheme and of its connection to Krawtchouk polynomials, it will be seen that the quantum walk on one graph of the scheme, namely the hypercube, precisely identifies with the 1-excitation dynamics of the spin chain associated to the Krawtchouk polynomials. Section 5 will offer a primer on bivariate Krawtchouk polynomials and their algebraic interpretation as matrix elements of the rotation group $SO(3)$ on the state vectors of the harmonic oscillator in three dimensions.
Furthermore, the ordered 2-Hamming scheme will be presented with some of its combinatorics. We shall find out that the adjacency matrices of this scheme are expressed as some bivariate Krawtchouk polynomials (of the Tratnik type) in two elementary matrices. Section 6 will be dedicated to the identification of a (weighted) graph in the ordered 2-Hamming scheme that has both FR and PST and which projects to a spin lattice with these two phenomena. We shall recap our findings in the last section. 

\section{Fractional revival (FR) and perfect state transfer (PST) in a one dimensional spin chain}
We shall here briefly review how the design of spin chains with FR and PST is carried out with the help of orthogonal polynomial theory.
Consider the $X\!X$ chain with the Hamiltonian on $(\mathbb{C}^2)^{N+1}$
\[
H = \frac{1}{2}\sum_{l=0}^{N-1} J_{l+1}(\sigma_l^x\sigma_{l+1}^x+\sigma_l^y\sigma_{l+1}^y)+\frac{1}{2}\sum_{l=0}^N B_l(\sigma_l^z+1),
\]
where $\sigma_l^x,\sigma_l^y,\sigma_l^z$ are the Pauli matrices acting at site $l$ as follows on the canonical basis $\{ \left|0\right>, \left| 1\right> \}$ for $\mathbb{C}^2$:
\begin{align*}
&\sigma^x \left|1\right> = \left| 0\right>,\quad \sigma^y \left|1\right> = i \left|0\right>, \quad  \sigma^z \left|1\right> = \left|1\right>,\\
&\sigma^x \left|0\right> = \left|1\right>,\quad  \sigma^y \left|0\right> = -i\left| 1\right>,\quad \sigma^z \left| 0 \right> =-\left|0\right>.
\end{align*}
It is not difficult to see that
\[
\left[ H, \sum_{l=0}^N \sigma_l^z\right]=0,
\]
in other words that the number of spins that are up is conserved. By convention we shall say that a spin is up if the state is described by the eigenvector of $\sigma^z$ that has eigenvalue $+1$. 
We shall use this property to restrict our considerations to the one-excitation subspace, that is to chain states where there is only one spin up.
A basis for this subspace will be provided by the following vectors in $\mathbb{C}^{N+1}$:
\[
\left| e_l\right) =(0,0,\ldots ,0,1,0,\ldots,0)\quad l=0,1,\ldots ,N,
\]
that have their single $1$ entry at the position corresponding to the site where the only spin up is located.
In view of the above conservation law, the Hamiltonian preserves the span of these vectors and is seen to act as follows on this basis:
\[
H \left| e_l\right) = J_{l+1}\left| e_{l+1}\right) +B_l\left| e_l\right) +J_l\left| e_{l-1}\right),\quad l=0,1,\ldots ,N
\]
with $J_0=J_{N+1}=0$.
In other words, in the occupation basis, the Hamiltonian in the one-excitation subspace takes the form of the following symmetric Jacobi matrix $J$
\[
J=\begin{pmatrix}
B_0 & J_1 &     & \\
J_1 & B_1 & J_2 & \\
    & J_2 & \ddots & \ddots \\
    &     &  \ddots    & B_{N-1} & J_N\\
    &     &        & J_N & B_N
\end{pmatrix}.
\]
Let us now consider the eigenvalue problem
\begin{equation}
H\left| s\right> = x_s\left|s\right>,\quad s=0,1,\ldots,N. \label{evp}
\end{equation}
and expand the eigenvectors in terms of the occupation basis vectors:
\[
\left| s\right> = \sum_{n=0}^N \sqrt{w_s}\chi_n(x_s)\left| e_n\right),\quad \chi_0(x_s)=1.
\] 
Since $H=J$ on the 1-excitation subspace, it is readily seen that the set $\{ \chi_n(x)\}_{n=0}^{N}$ is one of orthogonal polynomials whose orthogonality is given by
\[
\sum_{s=0}^N \chi_n(x_s)\chi_m(x_s)w_s=\delta_{mn}.
\]
From \eqref{evp}, one also finds the corresponding three-term recurrence relation
\[
x_s \chi_n(x_s)=J_{n+1}\chi_{n+1}(x_s)+B_n\chi_n(x_s)+J_n\chi_{n-1}(x_s).
\]
Since the matrix $(\sqrt{w_s}\chi_n(x_s))_{n,s=0}^N$ is an orthonormal matrix assuming that the eigenvectors are normalized, we have the inverse expansion
\begin{equation}\label{inv_expansion}
\left| e_n\right)=\sum_{s=0}^N \sqrt{w_s}\chi_n(x_s)\left|s\right>.
\end{equation}
We can now write down down how the FR and PST conditions translate on the spectra. From the FR condition namely,
\[
e^{-iTH}\left|e_0\right) =\mu \left|e_0\right)+\nu \left|e_N\right),\quad |\mu|^2+|\nu|^2=1,
\]
one finds with the help of \eqref{inv_expansion}
\begin{equation}
e^{-iTx_s}=\mu + \nu \chi_N(x_s)\label{fr1}
\end{equation}
since $\chi_0(x)=1$.
Let us begin with the analysis of PST which occurs when $\mu =0$ and $\nu = e^{i \phi}$ so that
\[
\chi_N(x_s)=e^{-i\phi }e^{-iTx_s},\quad \phi \in \mathbb{R},
\] 
which implies that $\chi_N(x_s)=\pm 1$ since $\chi_N(x_s)$ is real. 
A simple argument using the interlacing properties of zeros of orthogonal polynomials and the positivity of weight function yields
\begin{equation}
\chi_N(x_s)=(-1)^{N+s}\label{cond1}
\end{equation}
as a necessary condition for PST. It has also be shown \cite{VZ1} that the condition \eqref{cond1}
amounts in terms of the recurrence coefficients to the following mirror-symmetric (or persymmetric) requirement
\begin{equation}
J_{N-n+1}=J_n,\quad B_{N-n}=B_n. \label{mirror}
\end{equation}
We then proceed to find spin chains for which both \eqref{fr1} and \eqref{cond1} are satisfied, which means spin chains where both FR and PST take place. This is readily seen to imply
\begin{equation}
e^{-iTx_s}=e^{i\phi }( \cos \theta + i(-1)^{N+s}\sin \theta ). \label{fr2}
\end{equation}
It should be remarked that the case $\theta = \frac{\pi}{2}$ corresponds to PST since the amplitude $\mu $ is zero.

The problem at this point amounts to finding the parameters $J_n$ and $B_n$ given a spectral set that satisfy the FR requirements \eqref{fr2}.
We can attain this goal by constructing the associated monic orthogonal polynomials
\[
P_n(x)=\sqrt{J_1J_2\cdots J_n}\chi_n(x)
\]
whose recurrence coefficients will give $J_n$ and $B_n$.
Briefly, this can be done as follows. From the spectrum data $\{ x_s\}_{s=0}^N$, we can introduce the characteristic polynomials
\[
P_{N+1}(x)=(x-x_0)(x-x_1)\cdots (x-x_N),
\] 
which is orthogonal to all other $ P_n(x)\quad (n=0,1,\ldots,N)$. The condition \eqref{cond1} provides values for $P_N(x)\propto \chi_N(x)$ at $N+1$ points which fixes $P_N(x)$ by Lagrange interpolation.
Once two OPs $P_{N+1}(x),P_N(x)$ are known, all the others are obtained by the recurrence relation.
(For more details, the readers may consult \cite{GVZ1,VZ1})
\section{Para-Krawtchouk and Krawtchouk models}\label{sec:pkraw-kraw}
Let us see what this algorithm gives in an example.
Consider the following spectrum
\begin{equation}\label{bilattice}
x_s=\beta \left( s+\frac{1}{2}(\delta -1)(1-(-1)^s)-\frac{1}{2}(N-1+\delta )\right),\quad s=0,1,\ldots,N
\end{equation}
which can be viewed as the affine transformation of the superposition of 2 regular lattices of step 2 with spacing $\delta$. 
It can be checked that this set of spectral points satisfies the FR condition \eqref{fr2} with
\[
T=\frac{\pi }{\beta },\quad \theta =(-1)^N \frac{\pi }{2}\delta. 
\]
In order to have also PST, there must be a $t=T'$ such that 
\[
e^{-iT'x_s}=e^{i\phi }(-1)^{N+s},
\]
which requires $\delta =\frac{q}{p}$ with $p,q$ coprime integers (and $p$ also odd).
In this parametrization,  
\[
T'=qT.
\]
The spectrum \eqref{bilattice} will therefore correspond to a spin chain with both PST and FR. 
Using the reconstruction method for the Jacobi matrix that we described briefly at the end of the last section, we can obtain the chain specifications through the resulting recurrence coefficients of the associated polynomials. In the present case, when N is odd one finds:
\[
J_n=
\frac{\beta}{2}\sqrt{\frac{n(N+1-n)((N+1-2n)^2-\delta^2)}{(N-2n)(N-2n+2)}},\quad B_n=0.
\]
One observes that these couplings are indeed mirror symmetric.
Similar expressions are obtained for N odd. Remarkably, these explicit recurrence coefficients define orthogonal polynomials that had not really been studied.
We have called them para-Krawtchouk polynomials in particular because their orthogonality grid \eqref{bilattice} has resemblances with the spectrum of the parabosonic oscillator \cite{GVZ1,VZ2}.
Quite strikingly they emerge naturally when one looks for fractional revival \cite{BCB}.\par 

If we set $\delta =1$, the recurrence coefficients become
\begin{equation}\label{coeff_Krawtchouk}
J_n=\frac{\beta \sqrt{n(N+1-n)}}{2},\quad B_n=0,
\end{equation}
which are the coefficients of Krawtchouk polynomials which have the following explicit expression in terms of Gauss' hypergeometric series:
\[
K_n^N(x;p)=\mbox{}_2F_1\left( \begin{matrix} -n,-x\\ -N\end{matrix};\frac{1}{p}\right)=\sum_{k=0}^{N}\frac{(-n)_k(-x)_k}{k!(-N)_k}\left( \frac{1}{p}\right)^k,\quad (0<p<1)
\]
with $p=\frac{1}{2}$.
The corresponding spectrum of the Kratchouk polynomials are of course
\[
x_s= \beta \left( s-\frac{N}{2}\right)
\]
and 
\[
\theta = (-1)^N \frac{\pi }{2}.
\]
Therefore, only PST (not FR) can be observed in the spin chain associated with the Krawtchouk polynomials.
\section{Quantum walk on the hypercube}
We have seen that the Krawtchouk model \eqref{coeff_Krawtchouk} exhibits PST. It will be instructive to understand how this relates to quantum walks on the hypercube viewed as a graph of the (binary) Hamming scheme and to see how the Krawtchouk polynomials appear in this picture. This will allow us to review basic facts about a standard example of association schemes \cite{BCN}.

\subsection{A brief review of the Hamming scheme}\label{sec:hamming}
Recall that a graph $G=(V;E)$ is defined by a set of vertices $V$ and a set of edges $E$, which are 2-element subsets of $V$.
Let $|V|$ be a cardinality of $V$. The adjacency matrix $A$ of $G$ is a $|V|\times |V|$ matrix whose $(x,y)$ element $A_{xy}=\left< x| A |y\right>$ is given by the number of edges between vertices $x$ and $y$.
Now set $V=\{ 0,1\}^N$ which consists of $N$-tuples of $0$ and $1$. For these vertices, we can introduce the Hamming distance $d(x,y)$ between $x,y\in V$ which is the number of positions where $x$ and $y$ differ. Using the Hamming distance, we can also introduce the graphs $G_i~~(i=0,1,\ldots ,N)$ whose edges connect all pairs vertices with Hamming distance $i$. It should be remarked that $G_1$ is nothing but the $N$-dimensional hypercube. Let $A_i$ be the adjacency matrix of $G_i$ and $p_{ij}^k(=p_{ji}^k)$ be the intersection numbers which count the number of $z\in V$ such that
\[
d(x,z)=i,\quad d(y,z)=j \quad \textrm{if}\quad d(x,y)=k.
\]
The set of matrices $\{ A_i\}_{i=0}^N$ is known to satisfy the Bose-Mesner algebra:
\[
A_iA_j=\sum_{k=0}^N p_{ij}^k A_k,
\]
which is an essential feature of association schemes. The set of graphs $\{ G_i\}_{i=0}^N$ belongs to the one known as the binary Hamming scheme $\mathcal{H}(N,2)$. In this case, we have
\[
A_1 A_i=(i+1)A_{i+1}+(N-i+1)A_{i-1}
\]  
which implies $A_i=p_i(A_1)$, where $p_i(x)$ is a polynomial of degree $i$. One can further see that the polynomial $p_i(x)$ is the Krawtchouk polynomial \cite{St}:
\[
p_i(\lambda_s)=\binom{N}{i}K_i^N\left( s;\frac{1}{2}\right)\quad s=0,1,\ldots ,N
\]
with $\lambda_s =N-2s$.

\subsection{Projection of the quantum walk on the hypercube to the Krawtchouk model}\label{sec:projection}
Let us now explain how quantum walks on the hypercube can be projected to walks on a weighted path that can be identified with the 1-excitation dynamics of the Krawtchouk spin model \cite{CDDEKL}.\par
Let us consider the $N$-dimensional hypercube, i.e. the graph $G_1$ and denote its adjacency matrix by $A_1$. The unitary operator
\begin{equation}\label{evol}
U(t)=e^{-itA_1}
\end{equation}
defines a (continuous-time) quantum walk on the hypercube. 
We pick the vertex which corresponds to $(0)\equiv (0,0,\ldots, 0)$ as reference vertex and organize $V$ as a set of $N+1$ columns $V_n\quad (n=0,1,\ldots ,N)$ defined by
\[
V_n=\{ x\in V~|~d(0,x)=n\}
\]
whose cardinality is $|V_n|=k_n=\binom{N}{n}$. We denote the vertices in $V_n$ by $V_{n,m}\quad (m=1,2,\ldots,k_n)$. They all have $n$ $1$'s. It is not difficult to see that each $V_{n,m}$ is connected to the $ N-n$ elements of column $V_{n+1}$ obtained by converting a $0$ of $V_{n,m}$ to a $1$.\par
To the vertices $x\in V=\{ 0,1\}^N$,we shall associate orthonormalized vectors $\left| x\right> \in \mathbb{C}^{|V|}$ such that 
\[
\left< x| y\right> =\begin{cases} 1\quad \textrm{if}\quad d(x,y)=0\\ 0\quad \textrm{otherwise}\end{cases}\quad (x,y\in V)
\]
and introduce the linear span of the following column vectors:
\[
\left| \textrm{col}~n\right>=\frac{1}{\sqrt{k_n}}\sum_{m=1}^{k_n}\left| V_{n,m}\right>.
\]
The key observation here is that the evolution \eqref{evol} preserves column space because of
distance-regularity, i.e. each vertex in $V_n$ is connected to the same number of
vertices in $V_{n+1}$ and vice-versa. In light of this observation, it is possible to project quantum walks on the hypercube to quantum walk along the columns. We can realize this quotient by computing the matrix elements of $A_1$ between the column vectors. The non-zero elements are obtained as follows:
\begin{align*}
\left< \textrm{col}~n+1| A_1| \textrm{col}~n\right> &=\frac{1}{\sqrt{k_nk_{n+1}}}\sum_{m'=1}^{k_{n+1}}\sum_{m=1}^{k_n} \left< V_{n+1,m'}| A_1|V_{n,m}\right>\\
&=\frac{k_n(N-n)}{\sqrt{k_nk_{n+1}}}=\sqrt{(n+1)(N-n)}.
\end{align*}
By symmetry, we also have
\[
\left< \textrm{col}~n-1| A_1| \textrm{col}~n\right> = \left< \textrm{col}~n| A_1| \textrm{col}~n-1\right>= \sqrt{n(N-n+1)}.
\]
One thus finds that $A_1$ has the action
\[
A_1 \left| \textrm{col}~n\right> = J_{n+1}\left| \textrm{col}~n+1\right> +J_n \left| \textrm{col}~n-1\right>
\]
with $J_n=\sqrt{n(N-n+1)}$, which coincides with  that of $H$ on $\left| e_n\right)$ in the Krawtchouk model (up to a constant factor). It turns out that there is PST on the hypercube but no FR and this infers the same properties for the Krawtchouk model.
\section{Bivariate Krawtchouk polynomials}
In Section \ref{sec:pkraw-kraw} we have indicated how one-dimensional spin chains with PST and possibly FR could be engineered by identifying the couplings and local magnetic fields along the chain with the recurrence coefficients of suitable orthogonal polynomials. We have presented in some details models associated to the para-Krawtchouk and Krawtchouk polynomials in one variable. This suggests that spin lattices in dimensions higher than one could be constructed with the help of orthogonal polynomials in many variables. We shall focus on two dimensions in the following. With an eye to finding an example of a two-dimensional spin lattice with PST-like properties that extends the simplest system in 1D, we shall review results that concern bivariate Krawtchouk polynomials.\par
While univariate Krawtchouk polynomials are polynomials orthogonal with respect to binomial distribution function as follows:
\[
\sum_{k=0}^N \binom{N}{x}p^x(1-p)^{N-x}K_n(x)K_m(x)=h_n\delta_{m,n},
\]
bivariate Krawtchouk polynomials are orthogonal with respect to the trinomial distribution function $w(x,y)=\binom{N}{x,y}p^xq^y(1-p-q)^{N-x-y}$:
\begin{equation}\label{orthogonality:2-var}
\sum_{0\le x+y\le N}w(x,y)K_{m_1,n_1}(x,y)K_{m_2,n_2}(x,y)=h_{m_1,n_1}\delta_{m_1,m_2}\delta_{n_1,n_2}.
\end{equation}
It should be noted that the orthogonality condition \eqref{orthogonality:2-var} does not define bivariate Krawtchouk polynomials uniquely and hence there are variants.\par
Bivariate Krawtchouk polynomials of Tratnik are defined through the product of two univariate Krawtchouk polynomials \cite{Tr}:
\begin{equation}\label{Tratnik}
T_{m,n}^N(x,y)=\frac{(n-N)_m(x-N)_n}{(-N)_{m+n}}K_m^{N-n}(x;p)K_n^{N-x}\left( k,\frac{q}{1-p}\right).
\end{equation}
The Krawtchouk polynomials of Griffiths, originally introduced in \cite{DG,Gr} and rediscovered by Hoare and Rahman \cite{HR}, are usually defined as some specialization of the Aomoto-Gel'fand hypergeometric series \cite{GR}:
\begin{equation}\label{Griffith}
G_{m,n}^N(x,y)=\sum_{0\le i+j+k+l\le N }\frac{(-m)_{i+j}(-n)_{k+l}(-x)_{i+k}(-y)_{j+l}}{i!j!k!l!(-N)_{i+j+k+l}}u_1^iv_1^ju_2^kv_2^l
\end{equation} 
with
\begin{align}
\begin{split}
&pu_i+qv_i=1,\quad i=1,2,\\
&pu_1u_2+qv_1v_2=1.
\end{split}
\end{align}
It should be remarked here that the series \eqref{Griffith} reduces to \eqref{Tratnik} if we set
\begin{equation}\label{parameterization:Tratnik}
u_1=\frac{1}{p},\quad v_1=0,\quad u_2=1,\quad v_2=\frac{1-p}{q}.
\end{equation}
In other words, the Krawtchouk polynomials of Griffiths contains those of Tratnik as a special case. 
\subsection{Algebraic interpretation: $SO(3)$}
A group-theoretic interpretation of the multivariate Krawtchouk polynomials allows for a cogent derivation of many of their properties. It turns out that the Krawtchouk polynomials of Griffiths in $d$ variables can be interpreted as matrix elements of $SO(d+1)$ unitary representations \cite{GVZ1}. We shall give a brief review of this in the case $d=2$. \par
Let $a_i,a_i^+ (i=1,2,3)$ be operators of 3 independent oscillators with the action
\[
a_i\left| n_i\right> = \sqrt{n_i}\left| n_i-1\right>,\quad a_i^+\left| n_i\right> = \sqrt{n_i+1}\left| n_i+1\right>,\quad a_i\left| 0\right>=0.
\]
and let $\left| n_1,n_2,n_3\right>$ be the oscillator states defined by
\[
\left| n_1,n_2,n_3\right> =\left| n_1 \right> \otimes \left| n_2\right> \otimes \left| n_3 \right>.
\]
We fix $n_1+n_2+n_3=N$ and write 
\begin{equation}\label{basis}
\left| m,n\right>_N=\left| m,n,N-m-n\right>,\quad 0\le m+n\le N.
\end{equation}
Since the three-dimensional harmonic oscillator Hamiltonian $H=\sum_{i=1}^3 a_ia_i^+$ is invariant under $SU(3)$ and a fortiori under its $SO(3)$ subgroup, the eigensubspace of energy $N$ spanned by the (orthonormal) basis vectors $\left| m,n\right>_N$ forms a representation space for these groups. 
Let $R\in SO(3)$ and define its unitary representation $U=U(R)$ by
\begin{equation}\label{unitary}
U(R)a_iU^+(R)=\sum_{k=1}^3 R_{ki}a_k.
\end{equation}
The matrix elements of this unitary operator in the basis \eqref{basis} can be cast
in the form
\[
\tensor[_N]{\left< i,k| U(R)|m,n\right>}{_N}=w_{i,k;N} P_{m,n}^N(i,k)
\]
with $P_{0,0}^N(i,k)=1$ and $w_{i,k;N}=\tensor[_N]{\left< i,k| U(R)|0,0\right>}{_N}$. From the unitarity of $U$:
\[
\tensor[_N]{\left< m',n' | U^+ U| m,n\right>}{_N}= \sum_{0\le i+k\le N}\tensor[_N]{\left< m',n'| U^+ | i,k\right>}{_N }\tensor[_N]{\left< i,k|U|m,n\right>}{_N}=\delta_{m,m'}\delta_{n,n'},
\]  
it is straightforward to see that $\{ P_{m,n}\}_{m,n}$ have the orthogonality relation
\[
\sum_{0\le i+k\le N} w_{i,k;N}^2 P_{m,n}^N(i,k)P_{m',n'}^N(i,k)=\delta_{m,m'}\delta_{n,n'}.
\] 
The weight function $w_{i,k;N}^2$ can be computed directly as follows. From \eqref{unitary} and 
$\tensor[_{N-1}]{\left< i,k| U(R)a_1|0,0\right>}{_N}=0$, we see that
\begin{align*}
&\tensor[_{N-1}]{\left< i,k| Ua_1|0,0\right>}{_N}\\
&=\tensor[_{N-1}]{\left< i,k| Ua_1U^+Ua_1|0,0\right>}{_N}\\
&=R_{11}\sqrt{i+1} \tensor[_{N}]{\left< i+1,k| U|0,0\right>}{_N}+R_{21}\sqrt{k+1}\tensor[_{N}]{\left< i,k+1| U|0,0\right>}{_N}\\
&+R_{31}\sqrt{N-i-k}\tensor[_{N}]{\left< i,k| U|0,0\right>}{_N},
\end{align*}
which results in
\[
R_{11}\sqrt{i+1}w_{i+1,k;N}+R_{21}\sqrt{k+1}w_{i,k+1;N}+R_{31}\sqrt{N-i-k}w_{i,k;N}=0.
\]
Similarly using $\tensor[_{N-1}]{\left< i,k| U(R)a_2|0,0\right>}{_N}=0$, one finds 
\[
R_{12}\sqrt{i+1}w_{i+1,k;N}+R_{22}\sqrt{k+1}w_{i,k+1;N}+R_{32}\sqrt{N-i-k}w_{i,k;N}=0.
\]
It is not difficult to verify that
\[
w_{i,k;N}=C\frac{R_{13}^iR_{23}^kR_{33}^{N-i-k}}{\sqrt{i!k!(N-i-k)!}}
\]
is a solution to the above difference systems. The constant term $C$ is determined to be $C=\sqrt{N!}$ from the relation
\begin{align*}
1&= \tensor[_{N}]{\left< 0,0| U^+U|0,0\right>}{_N}\\
 &= \sum_{0\le i+k\le N} \tensor[_{N}]{\left< 0,0| U|i,k\right>}{_N} \tensor[_{N}]{\left< i,k| U|0,0\right>}{_N}=\sum_{0\le i+k\le N} w_{i,k;N}^2. 
\end{align*}
As a result, we have
\[
w_{i,k;N}=R_{13}^iR_{23}^kR_{33}^{N-i-k}\sqrt{\binom{N}{i,k}}
\]
and $P_{m,n}^N(i,k)$ are thus orthogonal with respect to trinomial distribution\\ $\binom{N}{i,k}p^iq^k(1-p-q)^{N-i-k}$ with
\begin{equation}\label{pqs}
p=R_{13}^2 ,\quad q=R_{23}^2.
\end{equation}
We can thus conclude that $\{ P_{m,n}^N(i,k)\}_{0\le m+n\le N}$ are (orthonormal) bivariate Krawtchouk polynomials. \par
The group theoretical interpretation enables us to derive several properties of $P_{m,n}^N(i,k)$. For instance, the relations
\begin{align*}
\tensor[_{N}]{\left< i,k| a_1^+a_1U|m,n\right>}{_N}=i \tensor[_{N}]{\left< i,k| U|m,n\right>}{_N}=\sum_{r,s=1}^3R_{r1}R_{s1}\tensor[_{N}]{\left< i,k| Ua_r^+a_s|m,n\right>}{_N},\\
\tensor[_{N}]{\left< i,k| a_2^+a_2U|m,n\right>}{_N}=i \tensor[_{N}]{\left< i,k| U|m,n\right>}{_N}=\sum_{r,s=1}^3R_{r2}R_{s2}\tensor[_{N}]{\left< i,k| Ua_r^+a_s|m,n\right>}{_N}
\end{align*}
yield the following two 7-term recurrence relations 
\begin{align}\label{7term-x}
\begin{split}
&iP_{m,n}^N(i,k)=[R_{11}^2 m+R_{12}^2 n+R_{13}^2(N-m-n)]P_{m,n}^N(i,k)\\
&+R_{11}R_{12}[\sqrt{m(n+1)}P_{m-1,n+1}^N(i,k)+\sqrt{n(m+1)}P_{m+1,n-1}^N(i,k)]\\
&+R_{11}R_{13}[\sqrt{m(N-m-n+1)}P_{m-1,n}^N(i,k)+\sqrt{(m+1)(N-m-n)}P_{m+1,n}^{N}(i,k)]\\
&+R_{12}R_{13}[\sqrt{n(N-m-n+1)}P_{m,n-1}^N(i,k)+\sqrt{(n+1)(N-m-n)}P_{m,n+1}^{N}(i,k)]
\end{split}
\end{align}
and
\begin{align}\label{7term-y}
\begin{split}
&kP_{m,n}^N(i,k)=[R_{21}^2 m+R_{22}^2 n+R_{23}^2(N-m-n)]P_{m,n}^N(i,k)\\
&+R_{21}R_{22}[\sqrt{m(n+1)}P_{m-1,n+1}^N(i,k)+\sqrt{n(m+1)}P_{m+1,n-1}^N(i,k)]\\
&+R_{21}R_{23}[\sqrt{m(N-m-n+1)}P_{m-1,n}^N(i,k)+\sqrt{(m+1)(N-m-n)}P_{m+1,n}^{N}(i,k)]\\
&+R_{22}R_{23}[\sqrt{n(N-m-n+1)}P_{m,n-1}^N(i,k)+\sqrt{(n+1)(N-m-n)}P_{m,n+1}^{N}(i,k)].
\end{split}
\end{align}
Furthermore, we can find that the $P_{m,n}^N(i,k)$ have the following explicit relation to the Krawtchouk polynomials of Griffiths \eqref{Griffith}:
\begin{align*}
P_{m,n}^N(i,k)=\sqrt{\binom{N}{m,n}}\left( \frac{R_{31}}{R_{33}}\right)^m\left( \frac{R_{32}}{R_{33}}\right)^nG_{m,n}^N(i,k)
\end{align*}
with 
\begin{align}\label{paramerization:rotation}
\begin{split}
u_1&=1-\frac{R_{11}R_{33}}{R_{13}R_{31}},\quad v_1=1-\frac{R_{21}R_{33}}{R_{23}R_{31}},\\ u_2&=1-\frac{R_{12}R_{33}}{R_{13}R_{32}},\quad v_2=1-\frac{R_{22}R_{33}}{R_{23}R_{32}}.
\end{split}
\end{align}
(For the details of the derivation, see \cite{GVZ1}.)
The Tratnik polynomials (of Krawtchouk type) are then seen to be the specialization of the Griffiths polynomials that corresponds to particular rotations given by the product of two rotation matrices about two orthogonal axes; for instance:  
\begin{align*}
R&=R_{yz}R_{xz}=\begin{pmatrix}
1 & 0 & 0\\
0 & \cos \theta & -\sin \theta \\
0 & \sin \theta & \cos \theta 
\end{pmatrix} \begin{pmatrix}
\cos \phi  & 0 & -\sin \phi \\
0 & 1 & 0 \\
\sin \phi  & 0 & \cos \phi  
\end{pmatrix}\\
&= \begin{pmatrix}
\cos \phi & 0 & -\sin \phi \\
\sin \theta \sin \phi & \cos \theta & \sin \theta \cos \phi\\
\cos \theta \sin \phi & -\sin \theta & \cos \theta \cos \phi
\end{pmatrix},
\end{align*}
from which $R_{12}=0$. The parametrization \eqref{pqs} and \eqref{paramerization:rotation} coincides with \eqref{parameterization:Tratnik}. The precise identification with the bivariate polynomials of Krawtchouk type defined in \eqref{Tratnik} involves a normalization factor given by:
\begin{equation}\label{Tratnik_orthonormal}
P_{i,j}^N(x,y)=\sqrt{\binom{N}{i,j}\tilde{p}^i\tilde{q}^j(1-p-q)^{-i-j}}T_{i,j}^N(x,y)
\end{equation}
with
\[
\tilde{p}=\frac{p(1-p-q)}{1-p},\quad \tilde{q}=\frac{q}{1-p}.
\]
It will be practical later to have handy the recurrence relations for the $T_{i,j}^N(x,y)$: 
\begin{align}\label{rec:Tratnik}
\begin{split}
xT_{i,j}^N(x,y)&=-p(N-i-j)[T_{i+1,j}^N(x,y)-T_{i,j}^N(x,y)]\\
&-(1-p)i[ T_{i-1,j}^N(x,y)-T_{i,j}^N(x,y)],\\
yT_{i,j}^N(x,y)&=\frac{pq}{1-p}(N-i-j)[T_{i+1,j}^N(x,y)-T_{i,j}^N(x,y)]\\
&-\frac{q}{1-p}(N-i-j)[ T_{i,j+1}^N(x,y)-T_{i,j}^N(x,y)]\\
&+qi[T_{i-1,j}^N(x,y)-T_{i,j}^N(x,y)]\\
&-\frac{p(1-p-q)}{1-p}j[T_{i+1,j-1}^N(x,y)-T_{i,j}^N(x,y)]\\
&-\frac{q}{1-p}i[T_{i-1,j+1}^N(x,y)-T_{i,j}^N(x,y)].
\end{split}
\end{align} 
For future reference also, let us record the following relation for the ``Hermitian'' Tratnik polynomials $P_{i,j}^N(x,y)$ which is obtained by combining the relations \eqref{rec:Tratnik} while mindful of \eqref{Tratnik_orthonormal}, with $p=\frac{1}{2}$ and $q=\frac{1}{4}$
\begin{align}\label{rec:Tratnik_orthonormal}
\begin{split}
&[\alpha (N-2x)+\beta (2N-2x-4y)] P_{i,j}^N(x,y)\\
&=\alpha j P_{i,j}^N(x,y) + \alpha \sqrt{(i+1)(N-i-j)}P_{i+1,j}^N(x,y)\\
&+\beta \sqrt{2(j+1)(N-i-j)}P_{i,j+1}^N(x,y)+\alpha \sqrt{i(N+1-i-j)}P_{i-1,j}^N(x,y)\\
&+\beta \sqrt{2j(N+1-i-j)}P_{i,j-1}^N(x,y)+\beta \sqrt{2i(j+1)}P_{i-1,j+1}^N(x,y)\\
&+\beta \sqrt{2(i+1)j}P_{i+1,j-1}^N(x,y).
\end{split}
\end{align}
The algebraic interpretation also allows us to obtain the generating function formula for $T_{i,j}^N(x,y)$:
\begin{align}\label{generating:Tratnik}
\begin{split}
&\sum_{0\le x+y\le N} \binom{N}{x,y}T_{i,j}^N(x,y)s^xt^y \\
& =(1+s+t)^{N-i-j}\left( 1+\frac{p-1}{p}s+t\right)^i\left( 1+\frac{p+q-1}{q}t\right)^j.
\end{split}
\end{align}

\subsection{Relationship to generalized Hamming scheme}
In section \ref{sec:hamming}, we have seen that univariate Krawtchouk polynomials naturally arise in the binary Hamming scheme $\mathcal{H}(N,2)$.
We shall here introduce the genralization of the Hamming scheme which is usually called the ordered $2$-Hamming scheme \cite{MS} and show that this scheme brings on the bivariate Krawtchouk polynomials of Tratnik.\par
Let $Q=\{ 0,1\}$ and consider the set $Q^{(N,2)}$ of vectors of dimension $2N$ over $Q$.
The vector $x\in Q^{(N,2)}$ will be presented by $2$-binary sequences of length $N$:
\[
x= (\bar{x}_1,\bar{x}_2,\ldots ,\bar{x}_N),\quad \bar{x}_j=(x_{j1},x_{j2})\in Q^2.
\]
For $x\in Q^{(N,2)}$, we can introduce the shape $e(x)$ by 
\begin{align*}
&e(x)=(e_1,e_2),\\
&e_1=\#\{ j\in \{1,2,\ldots,N\}~|~\bar{x}_j=(1,0)\},\\
&e_2=\#\{ j\in \{1,2,\ldots,N\}~|~\bar{x}_j=(0,1),(1,1)\} .
\end{align*}
For example, $x=((0,0),(1,0),(1,1),(0,1),(0,1))\in Q^{(5,2)}$ and $e(x)=(1,3)$.
We denote the set of the all shapes by
\[
E=\{ (e_1,e_2)\in (\mathbb{Z}_{\ge 0})^2~|~0\le e_1+e_2\le N\}.
\] 
We can now use shapes to establish relations between vertices. We shall say that two vertices $x,y \in Q^{N,2}$ are related under shape $e$:
\[
x\sim_e y\quad \textrm{if}\quad e((x-y \mod 2))=e.
\]
For example, let $x=((0,0),(1,0),(0,0))$ and $y=((1,1),(0,1),(1,0))$. Then $(x-y \mod 2)=((1,1),(1,1),(1,0))$ and $e((x-y \mod 2))=(1,2)$, we thus have $x\sim_{(1,2)}y$.
We can then introduce the graph $G_e$ associated with the shape $e$ as the one where all pairs of vertices $(v_x,v_y)$ in $\{ v_x~|~x\in Q^{(N,2)}\}$ are linked if $v_x \sim_e v_y$.
The adjacency matrix $A_e$ of the graph $G_e$ is given by
\[
\left< x|A_e|y\right> = \begin{cases} 1\quad (x\sim_e y)\\ 0\quad (\textrm{otherwise})\end{cases}.
\]
The set of the adjacency matrices $\mathcal{A}=\{ A_e~|~e\in E\}$ satisfy the Bose-Mesner algebra 
\[
A_{(i,j)}A_{(k,l)}=\sum_{0\le m+n\le N} \alpha_{(i,j),(k,l)}^{(m,n)} A_{(m,n)}
\]
and thus defines an association scheme called the ordered $2$-Hamming scheme.
The intersection numbers are here defined by 
\[
\alpha_{(i,j),(k,l)}^{(m,n)}=\#\{ z\in Q^{(N,2)}~|~x\sim _{(i,j)}z,y\sim_{(k,l)}z,x\sim_{(m,n)}y\}.
\] 
and it is found in particular that 
\begin{align}\label{bm:ordered-hamming}
\begin{split}
A_{(1,0)}A_{(i,j)}&=(N+1-i-j)A_{(i-1,j)}+jA_{(i,j)}+(i+1)A_{(i+1,j)},\\
A_{(0,1)}A_{(i,j)}&=2(N+1-i-j)A_{(i,j-1)}+2(i+1)A_{(i+1,j-1)}\\
&+(j+1)A_{(i-1,j+1)}+(j+1)A_{(i,j+1)}.
\end{split}
\end{align}
The derivation of these relations is not difficult. For instance, we can compute $\alpha_{(1,0),(i,j)}^{(i-1,j)}$ as follows. Let us count the number of $z\in Q^{(N,2)}$ such that
\[
e((x-z\mod 2))=(1,0),\quad e((y-z\mod 2))=(i,j)
\]
if $e((x-y\mod 2))=((i-1,j))$. Take 
\[
x=((1,0),\ldots,(1,0),(0,1),\ldots ,(0,1),(0,0),\ldots ,(0,0))
\]
with $i-1$ $(1,0)$s, $j$ $(0,1)$s and $N+1-i-j$ $(0,0)$s and $y=((0,0),\ldots ,(0,0))$. In order to get a $z$, a $(0,0)$ should be converted into a $(1,0)$ and there are $N+1-i-j$ ways to do that. We thus conclude that
\[
\alpha_{(1,0),(i,j)}^{(i-1,j)}=N+1-i-j.
\]
The other intersection numbers can be computed in the same manner.
Quite interestingly, the relations \eqref{bm:ordered-hamming} coincide with the recurrence relations \eqref{rec:Tratnik} for the bivariate Krawtchouk polynomials of Tratnik with $p=\frac{1}{2},q=\frac{1}{4}$.
Therefore, the bivariate Krawtchouk polynomials of Tratnik arise in the ordered $2$-Hamming scheme as the univariate Krawtchouk polynomials do in the binary Hamming scheme $\mathcal{H}(N,2)$.
\section{FR and PST in two-dimensional spin lattices}
We have seen in section \ref{sec:projection} that the quantum walk on the hypercube in the Hamming scheme leads to a 1-dimensional spin model with PST. We shall here consider the graph $G_{\alpha ,\beta }$ in the ordered $2$-Hamming scheme corresponding to the weighted adjacency matrix:
\[
A_{\alpha ,\beta }=\alpha A_{(1,0)}+\beta A_{(0,1)}
\] 
and examine the quantum walks this matrix generates. Thanks to the distance regularity of the graph, we can project the quantum walk on the graph $G_{\alpha ,\beta }$ to walks on a two-dimensional regular lattice of triangular shape in the same fashion as in section \ref{sec:projection}. \par 
Let $(0)\equiv ((0,0),(0,0),\ldots ,(0,0))$ be a reference vertex and organize $V$ as the set of $\binom{N+1}{2}$ column $V_{i,j}$ defined by
\[
V_{ij}=\{ x\in V~|~e(x)=(i,j)\}\quad 0\le i+j\le N.
\]
The cardinality of $V_{i,j}$ is given by
\[
k_{i,j}=|V_{i,j}|=\binom{N}{i,j}2^j.
\]
Let us label the vertices in column $V_{i,j}$ by $V_{(i,j),k},~~k = 1,2,\ldots ,k_{i,j}$.
Under the relation corresponding to shape $(1,0)$, each $V_{(i,j),k}$ in $V_{i,j}$ is connected to $N-i-j$ vertices in $V_{i+1,j}$ since exchanging a $(0,0)$ for a $(1,0)$ in $V_{(i,j),k}$ gives a vertex in $V_{i+1,j}$ and this can be done in $N-i-j$ ways.
It is not difficult to see similarly that $V_{(i,j),k}$ is connected to $j$ vertices in $V_{i,j}$. For the relation corresponding to the shape $(0,1)$, we can also see that each $V_{(i,j),k}$ in $V_{i,j}$ is connected to $2(N-i-j)$ vertices in $V_{i,j+1}$ and to $j$ vertices in $V_{i+1,j-1}$.\par
Consider now the column space taken to be the linear span of the column vectors
given by
\[
\left| \textrm{col}~i,j\right> = \frac{1}{\sqrt{k_{i,j}}}\sum_{k=1}^{k_{i,j}}\left| V_{(i,j),k}\right>.
\]
Distance regularity assures that $A_{(1,0)}$ and $A_{(0,1)}$ preserve column space and allows to project from the quantum walks on $G_{\alpha ,\beta }$ to the simplex labelling the columns.
To that end, we compute the matrix elements of $A_{(1,0)}$ and $A_{(0,1)}$ between the column vectors in the same manner as for the hypercube.
One finds:
\begin{align*}
\left< \textrm{col}~i+1,j|A_{(1,0)}|\textrm{col}~i,j\right>&=\sqrt{(i+1)(N-i-j)},\\
\left< \textrm{col}~i,j|A_{(1,0)}|\textrm{col}~i,j\right>&=j,\\
\left< \textrm{col}~i,j+1|A_{(1,0)}|\textrm{col}~i,j\right>&=\sqrt{2(j+1)(N-i-j)},\\
\left< \textrm{col}~i+1,j-1|A_{(1,0)}|\textrm{col}~i,j\right>&=\sqrt{2(i+1)j}.
\end{align*}
One thus observes that quantum walks on $G_{\alpha ,\beta }$ are in correspondence with the 1-excitation dynamics of the spin lattice of triangular shape with Hamiltonian \cite{MTV}
\begin{align}
\begin{split}\label{hamiltonian:ohs}
H=&\sum_{0\le i+j\le N} 
\alpha \sqrt{(i+1)(N-i-j)}\frac{\sigma_{i,j}^x \sigma_{i+1,j}^x+\sigma_{i,j}^y\sigma_{i+1,j}^y}{2}\\
&\quad \quad +\beta \sqrt{2(j+1)(N-i-j)}\frac{\sigma_{i,j}^x \sigma_{i,j+1}^x+\sigma_{i,j}^y\sigma_{i,j+1}^y}{2}\\
&\quad \quad +\beta \sqrt{2(i+1)j}\frac{\sigma_{i,j}^x \sigma_{i+1,j-1}^x+\sigma_{i,j}^y\sigma_{i-1,j+1}^y}{2}+ \alpha j \frac{1+\sigma_{i,j}^z}{2}.
\end{split}
\end{align}
Indeed, on the subspace spanned by the 1-excitation basis vectors $\left| e_{i,j}\right)=E_{i,j}$ with $E_{i,j}$ the $(N+1)\times (N+1)$ matrix with $1$ in the $(i,j)$ entry and zeros everywhere else, we see that
\begin{align}\label{hamiltonian-action}
\begin{split}
H\left| e_{i,j}\right) &= \alpha \sqrt{(i+1)(N-i-j)}\left| e_{i+1,j}\right) + \beta \sqrt{2(j+1)(N-i-j)}\left| e_{i,j+1}\right)\\
&+ \alpha \sqrt{i(N+1-i-j)}\left| e_{i-1,j}\right)  + \beta \sqrt{2j(N+1-i-j)}\left| e_{i,j-1}\right) \\
&+ \beta \sqrt{2(i+1)j}\left| e_{i+1,j-1}\right)+ \beta \sqrt{2i(j+1)} \left| e_{i-1,j+1}\right)+ \alpha j\left| e_{i,j}\right),
\end{split}
\end{align}
which corresponds to $[\alpha A_{(1,0)}+\beta A_{(0,1)}]\left| \textrm{col}~i,j\right>$
since the coefficients of the relation \eqref{hamiltonian-action} coincide with those of \eqref{rec:Tratnik_orthonormal}.
On the subspace spanned by the 1-excitation basis, $H$ can hence be diagonalized by the Hermitian Tratnik polynomials \eqref{Tratnik_orthonormal} with $p=\frac{1}{2}$ and $q=\frac{1}{4}$ and the corresponding eigenvalues and eigenvectors are given by
\begin{align*}
\lambda_{x,y}&=\alpha (N-2x)+\beta (2N-2x-4y),\\
\left|x,y\right>&=\sum_{0\le i+j\le N} w_{x,y;N}P_{i,j}^N(x,y)\left|e_{i,j}\right),\quad 0\le x+y\le N  
\end{align*}
respectively.\par
\begin{figure}[htbp]
  \begin{center}
    \begin{tabular}{c}

      \begin{minipage}{0.33\hsize}
        \begin{center}
          \includegraphics[width=3cm]{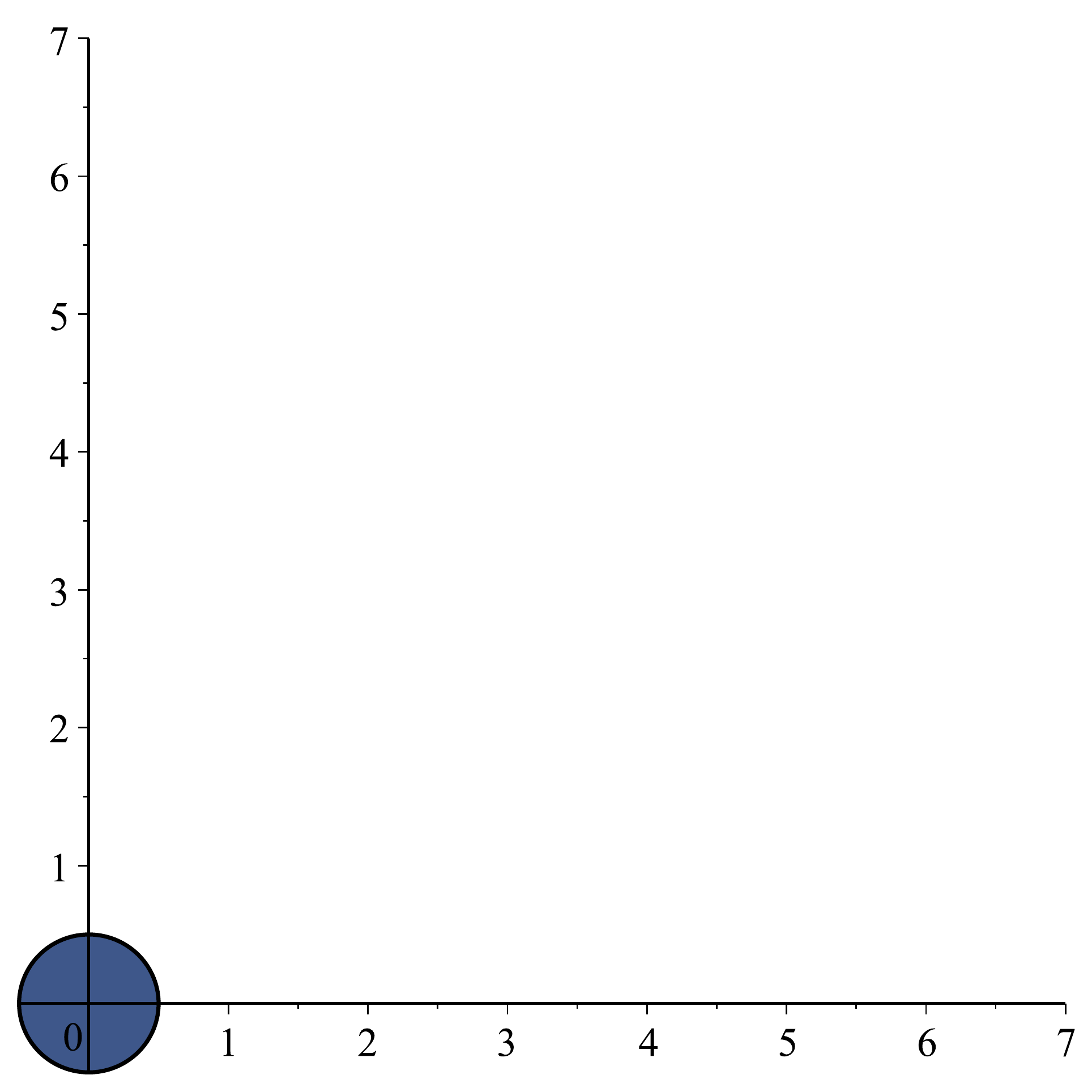}
         \hspace{1.6cm} $t=0$
        \end{center}
      \end{minipage}

      \begin{minipage}{0.33\hsize}
        \begin{center}
          \includegraphics[width=3cm]{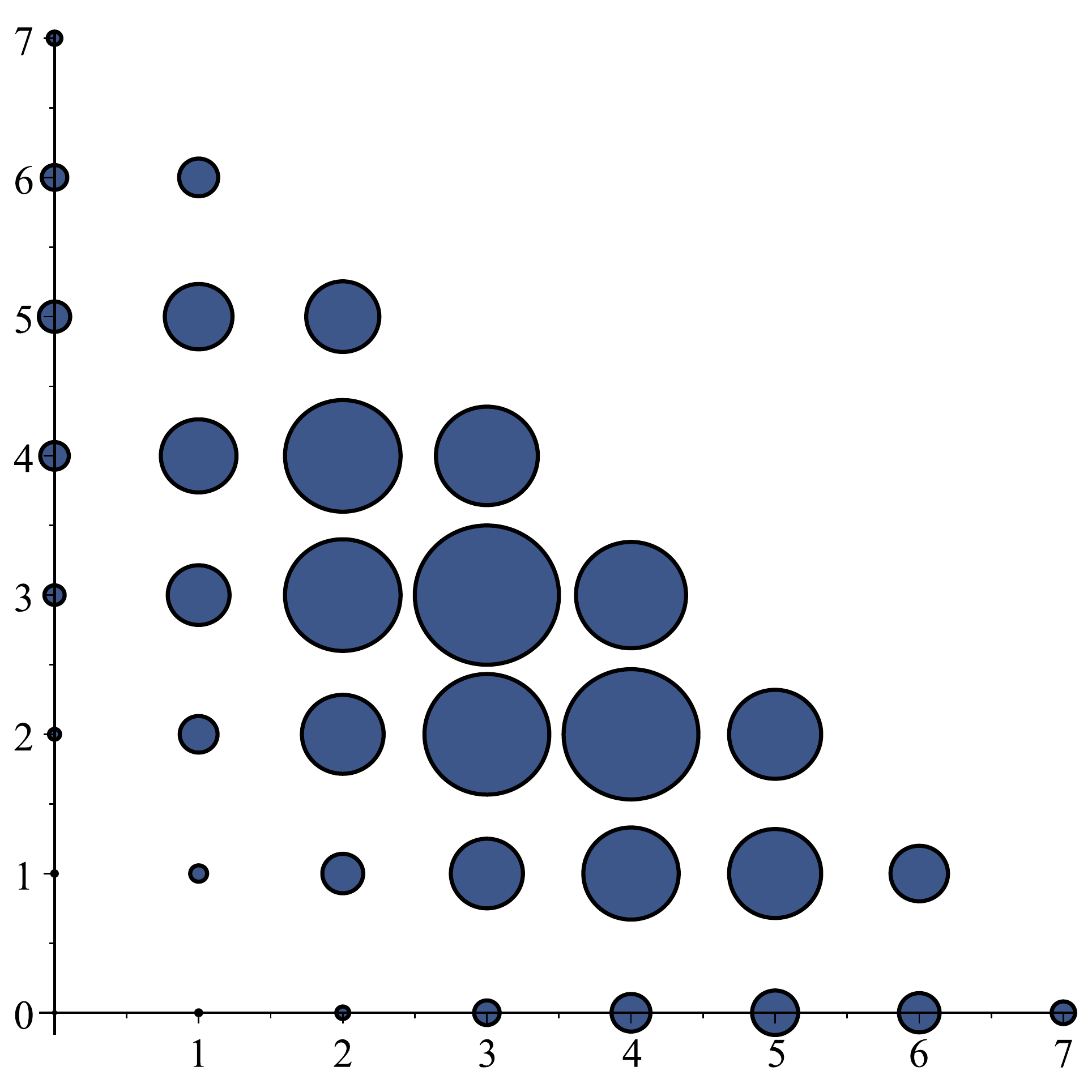}
                   \hspace{1.6cm} $t=\frac{\pi }{6}$
        \end{center}
      \end{minipage}

      \begin{minipage}{0.33\hsize}
        \begin{center}
          \includegraphics[width=3cm]{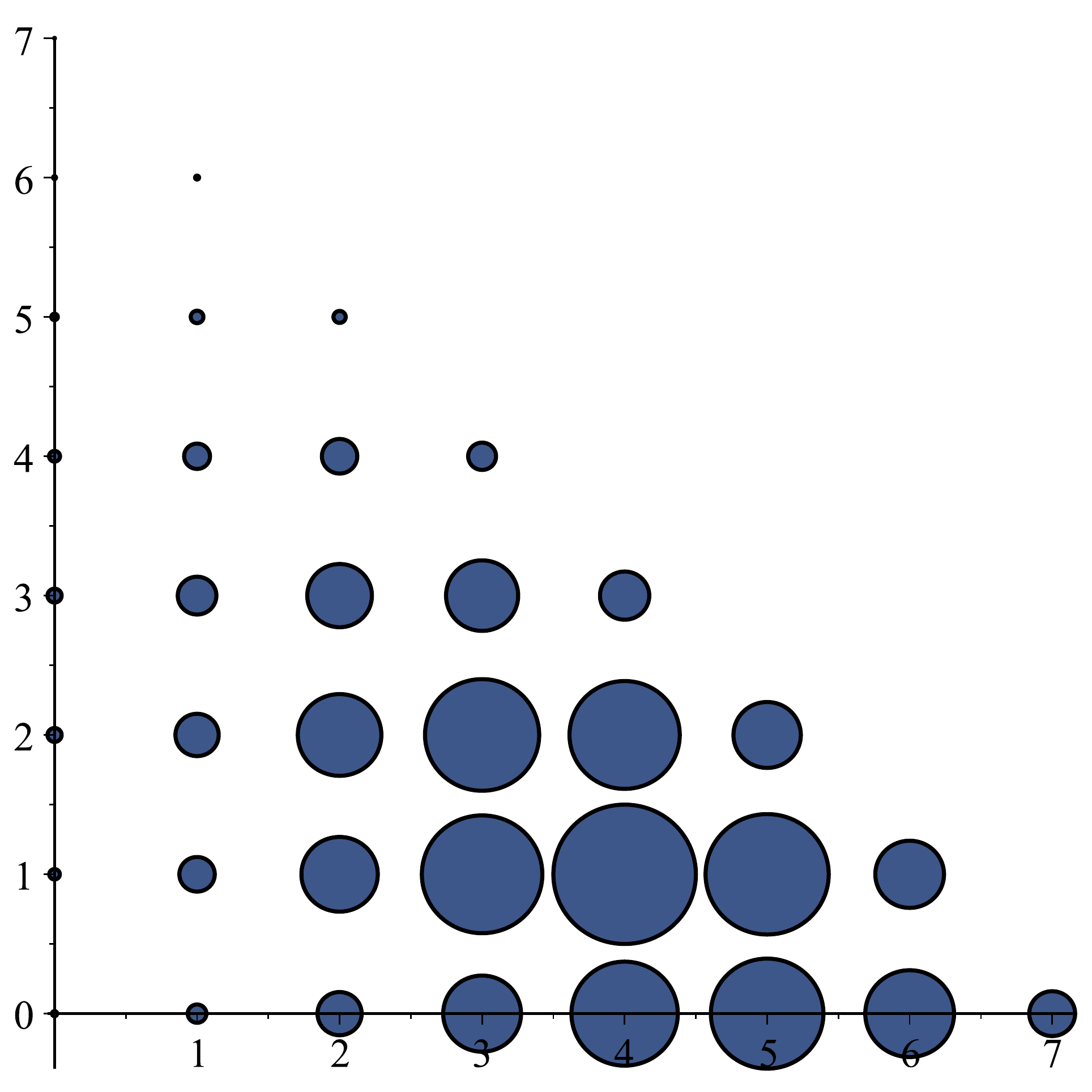}
                   \hspace{1.6cm} $t=\frac{\pi}{5}$
        \end{center}
      \end{minipage}
    \end{tabular}
        \begin{tabular}{c}

      \begin{minipage}{0.33\hsize}
        \begin{center}
          \includegraphics[width=3cm]{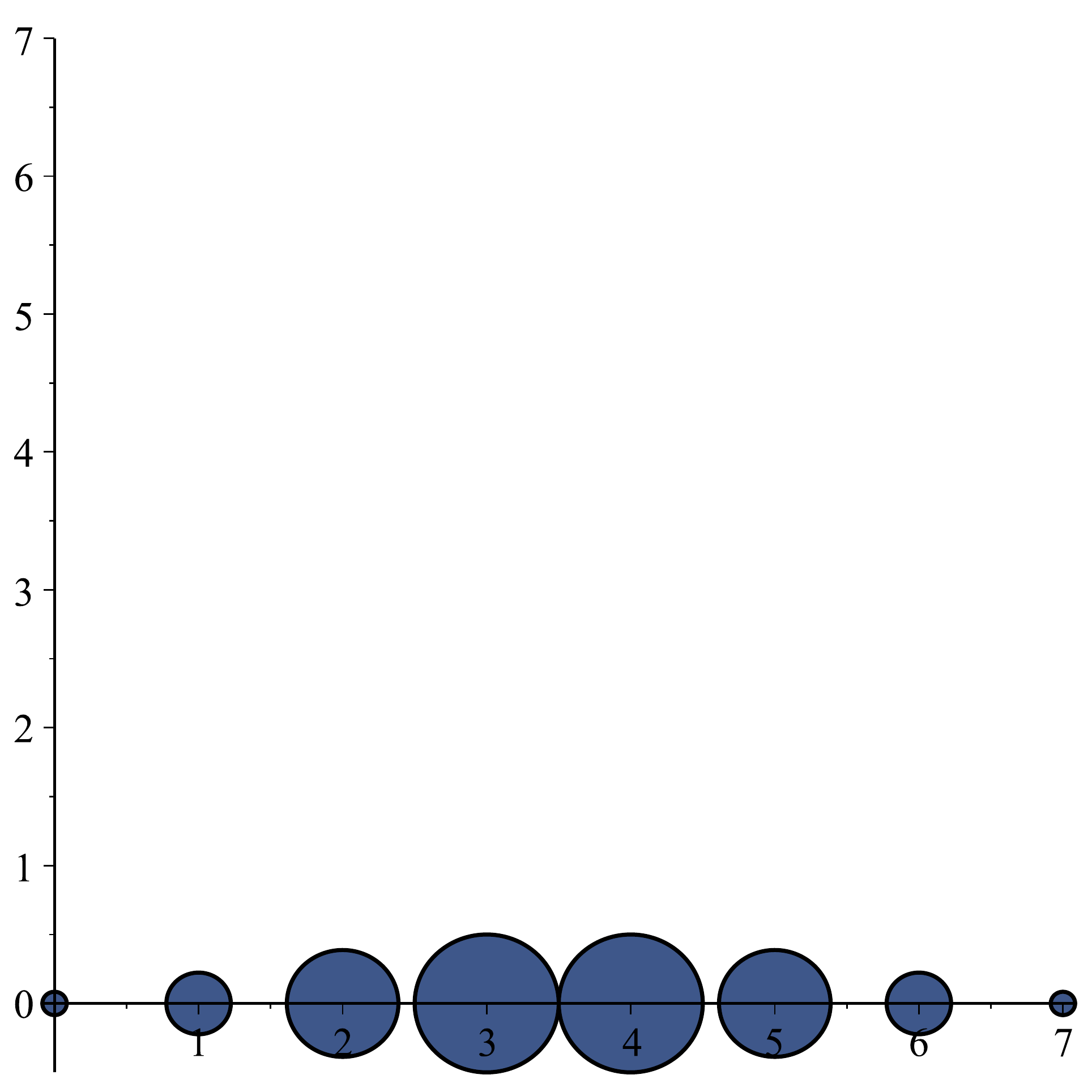}
                   \hspace{1.6cm} $t=\frac{\pi}{4}$
        \end{center}
      \end{minipage}

      \begin{minipage}{0.33\hsize}
        \begin{center}
          \includegraphics[width=3cm]{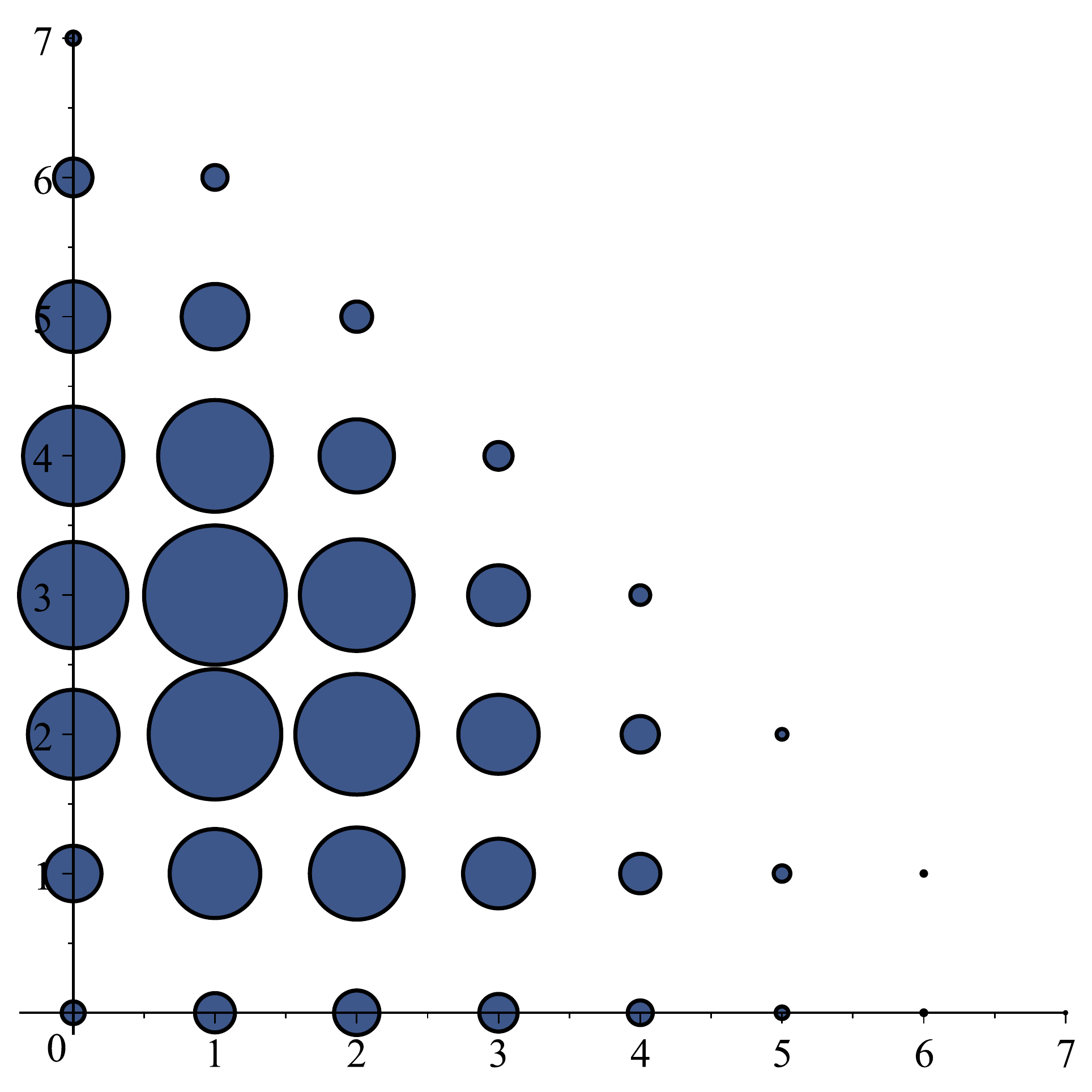}
                   \hspace{1.6cm} $t=\frac{\pi}{3}$
        \end{center}
      \end{minipage}

      \begin{minipage}{0.3\hsize}
        \begin{center}
          \includegraphics[width=3cm,clip]{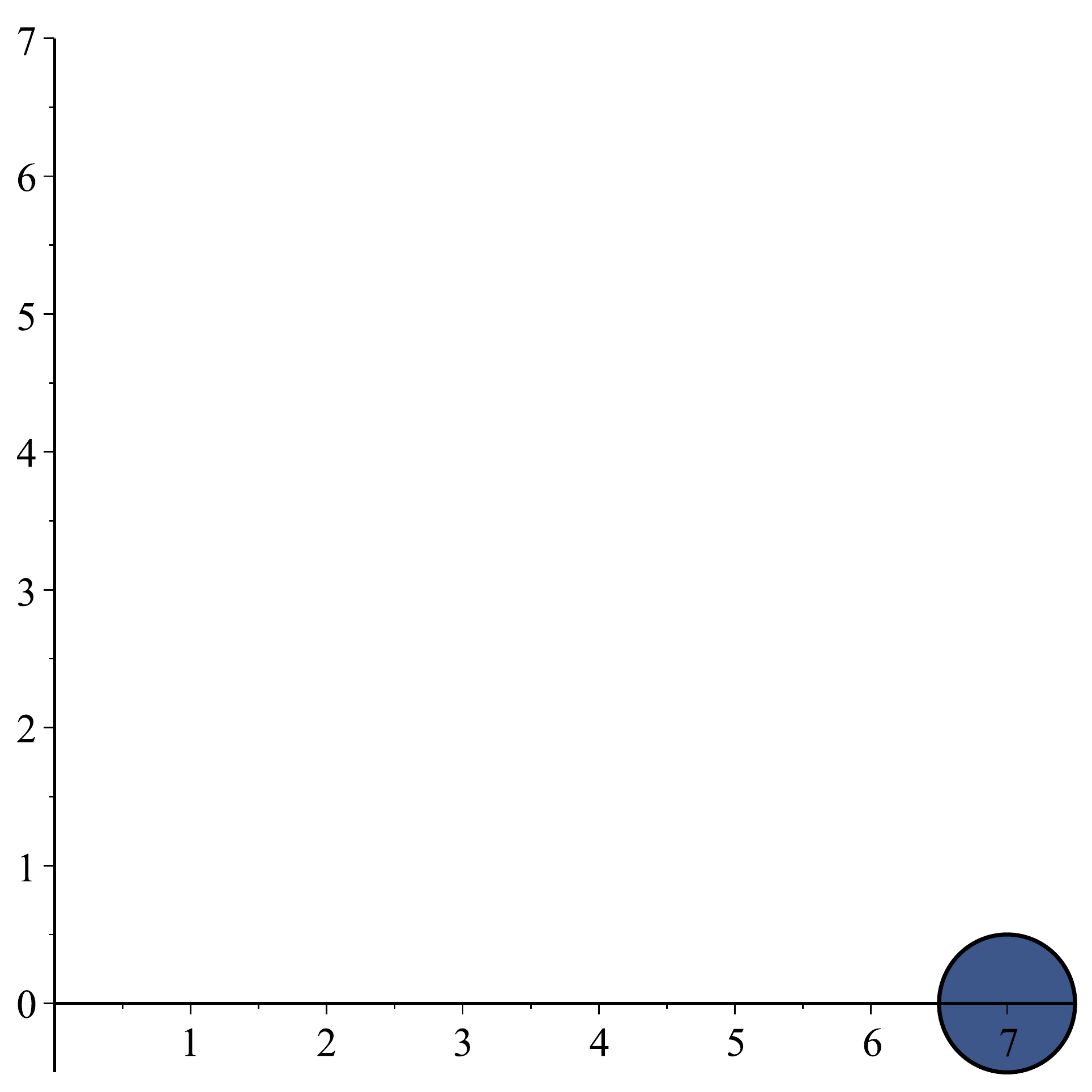}
        \end{center}
                 \hspace{1.6cm} $t=\frac{\pi}{2}$
      \end{minipage}
    \end{tabular}
    \caption{The transition amplitude $|f_{i,j}(t)|$ for $A_{(1,0)}+2A_{(0,1)}$ when $N=7$. The areas of the circles
are proportional to $|f_{(i,j)}(t)|$ at the given lattice point $(i,j)$. PST occurs at $\frac{\pi}{2}$ and FR on the set of sites $i=0,1,\cdots ,N$ and $j=0$ occurs at $t=\frac{\pi}{4}$.}
    \label{fig:a1b2}
  \end{center}
\end{figure}

 
Let us now examine the evolution of a single qubit located at the apex $(0,0)$ under the dynamics of this spin lattice.
The amplitude for finding that qubit at the site $(k,l)$ at time $t$ is given by:
\[
f_{(k,l)}(t)=\left< e_{k,l}| e^{-itH}|e_{0,0}\right>.
\]
This transition amplitude can be computed with the help of the generating function formula \eqref{generating:Tratnik} as follows:
\begin{align*}
f_{(k,l)}(t)&=\left< e_{k,l}| e^{-itH}|e_{0,0}\right>\\
&=\sum_{0\le x+y\le N} \left< e_{k,l}|x,y\right> e^{-it\lambda_{x,y}}\left< x,y|e_{0,0}\right>\\
&=\sum_{0\le x+y\le N} \binom{N}{x,y}\left( \frac{1}{2}\right)^x \left( \frac{1}{4}\right)^y \left( \frac{1}{4}\right)^{N-x-y}P_{0,0}^N(x,y)P_{k,l}^N(x,y)e^{-it\lambda _{x,y}}\\
&= e^{-iN(\alpha +2\beta )t}\frac{\sqrt{2^l}}{4^N}\sqrt{\binom{N}{k,l}}(1+2z_1+z_2)^{N-k-l}(1-2z_1+z_2)^k(1-z_2)^{l}
\end{align*}
with $z_1=e^{2i(\alpha +\beta )t}$ and $z_2=e^{4i\beta t}$.
In order to achieve transfer only to sites $(i,j)$ with $i+j=N$, we must have for some $t=T$
\[
1+2z_1+z_2=0.
\]
Since $|z_1|=|z_2|=1$, this last relation implies that
\begin{equation}\label{paramerization_for_pst}
z_2=1 ,\quad z_1=-1.
\end{equation}
From the final expression for the amplitude $f_{k,l}(t)$, we see that $z_2=1$ requires that $j=0$ at $t=T$ and hence 
\[
|f_{(i,j)}(T)|=\begin{cases}1\quad (i,j)=(N,0)\\ 0\quad (\textrm{otherwise})
\end{cases}.
\]
This is simply PST between $(0,0)$ and $(N,0)$.
The condition \eqref{paramerization_for_pst} for PST can be realized in different ways. One simple instance is  
\[
\alpha =1,\quad \beta =2,\quad T=\frac{\pi}{2}.
\] 
It should be remarked here that in the case $\alpha =1$ and $\beta =2$, we see that at $t=\frac{\pi}{4}$
\[
z_2=1,\quad z_1\ne -1,
\]
which implies
\[
\left| f_{(i,j)}\left( \frac{\pi}{4}\right)\right|=0,\quad j\ne 0.
\]
In other words, at $t=\frac{\pi}{4}$ FR occurs on the sites $(i,0),~i=0,1,\ldots, N$. This is depicted on the Fig. \ref{fig:a1b2} where $N=7$.

\section{Concluding Remarks}
Summing up, we have seen how PST and FR can occur on weighted paths associated to spin chains with non-uniform couplings prescribed by orthogonal polynomials. 
We have observed in particular that the Krawtchouk spin chain model is related with quantum walks on the hypercube.
This connection is underscored by the occurrence of Krawtchouk polynomials as eigenfunctions and as matrix eigenvalues of the Hamming scheme.
We have reviewed the algebraic interpretation of two-variable generalizations of the Krawtchouk polynomials and discussed the ordered $2$-Hamming scheme which features the bivariate Krawthouk polynomials of Tratnik.
This connection enabled us to introduce a new two-dimensional spin lattice model in two dimensions where PST and FR can be found. It should be mentioned that a spin lattice associated with the more general bivariate Krawtchouk polynomials, i.e. those of Griffiths, was considered \cite{MTVZ,Po} although PST was not observed in this  model.
We trust this lecture has illustrated the role that the theory of orthogonal polynomials can play in the analysis of quantum information tasks.


\end{document}